**Preprint Version**
**ICICML 2024**

# A Consolidated Volatility Prediction with Back Propagation Neural Network and Genetic Algorithm

---

2024

# A Consolidated Volatility Prediction with Back Propagation Neural Network and Genetic Algorithm

$Zong\ Ke^{1st*}$
Faculty of Science
National University of Singapore, Singapore 119077
a0129009@u.nus.edu

Jingyu Xu[2nd]
Northern Arizona University, Arizona 86011, USA
jyxu01@outlook.com

$Zizhou\ Zhang^{3rd}$
University of Illinois Urbana-Champaign, IL 61801, USA
zhangzizhou_2000@outlook.com

Yu Cheng[4th]
The Fu Foundation School of Engineering and Applied Science,
Columbia University, New York, NY 10027, USA
yucheng576@gmail.com

Wenjun Wu[5th]
The Grainger College of Engineering
University of Illinois Urbana-Champaign, IL 61801, USA
wenjun5@illinois.edu

* *Corresponding author*: Zong Ke
a0129009@u.nus.edu

***Abstract*—This paper provides a unique approach with AI algorithms to predict emerging stock markets volatility. Traditionally, stock volatility is derived from historical volatility, Monte Carlo simulation and implied volatility as well. In this paper, the writer designs a consolidated model with back -propagation neural network and genetic algorithm to predict future volatility of emerging stock markets and found that the results are quite accurate with low errors.**

## I. INTRODUCTION

Applying artificial intelligence algorithms to the field of financial investment is the latest hot research topic, especially the use of neural networks and other deep learning algorithms, which has pushed financial investment to the forefront. This paper will use a genetic algorithm, optimized BP neural network (hereinafter called GA-BP), to predict the volatility of the stock market.

As one of most popular neural networks in industry, BP network and its variations play a vital role in forward network, reflecting the basic influence of artificial neural networks. There are several areas that BP network is frequently leveraged.

In the 1970s, inspired by the evolution laws of organisms in nature, John Holland proposed genetic algorithm (GA). This model takes advantage of biological evolution to simulates the biological selection and genetic evolution.

Volatility prediction is of great significance in the financial market and is mainly used in quantitative trading, risk management, asset allocation and other aspects [1]. Volatility reflects the degree of volatility of financial asset prices and is a key indicator for measuring the uncertainty and risk level of asset returns. Volatility is the degree of volatility of financial asset prices and is used to reflect the uncertainty of asset returns. The higher the volatility is, the more violent the fluctuation of asset prices is, and the greater the uncertainty of returns is. On the contrary, the lower the volatility, the smoother the price fluctuation and the greater the certainty of returns. Volatility can be divided into the following categories. Historical volatility is calculated based on the fluctuation of investment performance during past periods. Implied volatility refers to the volatility obtained by reverse calculation through the option pricing model, reflecting the market's expectations of future volatility. Expected volatility is a forecast of future volatility based on historical data and market conditions.

## II. LITERATURE REVIEW

Existing researches have explored into different approaches for volatility prediction, evolving from traditional models such as autoregressive regression [2] and GARCH family models [3] to artificial intelligence models covering SVM, Lasso, Random Forest, XGBoost and deep learning. These models apply the latest progresses of machine learning in Finance. And each of them has their own advantages and disadvantages. Recently, hybrid models are proposed to further improve forecasting performance.

Researchers proposed two complicated models using the artificial neural network (ANN) and E-GARCH models to derive the volatility of the S&P500 index returns in order to improve the predictive ability of the GARCH family of models [4]. The fitting values of the EGARCH model with the best predictive effect were considered as input variables of a neural network. In the first hybrid model, the input variables only included all the explanatory variables' historical data and the volatility estimation values of the EGARCH model. In the second hybrid model, the input was increased by adding a simulated generated synthetic sequence, thus introducing the self-correlation structure captured by the EGARCH model and maintaining the advantages of the EGARCH. An innovative hybrid model that synergistically combines the predictive power of the random forest and the Lasso has a higher prediction accuracy of stock market volatility based on monthly macroeconomic variables. Bu et al. (2019) put forward a deep CNN with locality and sparsity constraints for texture classification [5]. Gao et al. (2019) have tried learning a robust representation via a deep CNN on symmetric positive definite manifolds [6]. Sui et al. (2024) proposed an ensemble approach, which combines deep learning and time series models, and proved to outperform single model [7]. Gong et al. (2024) designed a CNN-LSTM hybrid model [8]. Dong et al. (2024) integrated reinforcement learning into graph neural networks [9]. Dan et al. (2024) proposed a multi-view stereo reconstruction method and an improved you Only Look Once Algorithm based on deep learning [10-11]. Weng et al. (2024) explored the application of AI in finance [12]. Furthermore, Liu et al. (2024) applied an ANN and LSTM-based ensemble model in stock market prediction [13]. Li et al. (2024) incorporated contrastive learning to traditional deep learning to improve the portfolio performance of cryptocurrency and US treasuries [14].

Traditional models in existing researches have multiple strict constraints, such as positive definite variance, which will not hold in extreme scenario. Meanwhile, they have a high requirement of the quality of input data, whose stability and accuracy will be negatively influenced in case of outliers and noises. In contrast, strong non-linear mapping capability enables BP neural network to process nonlinear problems [15-17]. Meanwhile, global search and regularization term of GA avoids local optima and overfitting[18]. Therefore, in this paper, considering both emerging market conditions and the clustering effect of volatility data, we will use a combination of genetic algorithms and BP neural networks to derive a model for predicting the volatility of emerging markets[19-21].

## III. MODEL

### A. Data

There are many indices globally, such as S&P 500, DAX, FTSE 100, HSI, KOSPI and NIKKEI 225 and on. And most of financial theories are derived with them or with related underlying[22-23]. To build a model with broader generalization ability, emerging markets indices, especially those in China, will be first considered. And the CSI 300 Index is one of the most typical indices jointly released by the Shenzhen and Shanghai stock exchanges on April 8, 2005, reflecting the composition and performance of the China big-cap and mid-cap firms. We selected the sample period as the past ten years, from May 6, 2014, to May 26, 2024. Index data were fetched from the Wind database (the Wind Economic Database is one of the most widely adopted database in the financial industry in China, whose clients cover regulatory agencies, investment banks, asset management firms and universities)[24]. Preprocessing has been conducted before further analysis, including treatment of missing values and outliers using interpolation and normalization respectively.

First, using the price2ret function and the daily closing price data of the CSI 300 Index, we calculated the logarithmic returns, and the descriptive statistical results are shown in Table 1:

Table 1 descriptive statistical analysis of CSI 300 series

| | |
|---|---|
| Obs | 2783 |
| Mean | 0.0005 |
| Max | 0.1300 |
| Min | −0.1831 |
| S.D. | 0.0308 |
| Skewness | −0.6632 |
| Excess Kurtosis | 4.2802 |
| $Q^2(10)^a$ | 31.3538 |
| ARCH test$(10)^b$ | 311.2793 |

From above, the data exhibits a bimodal and leptokurtic distribution. The results of the LBQ test and the ARCH test indicate that data is auto-correlated.

In this paper, we selected both endogenous and exogenous variables. For the Chinese market, endogenous explanatory variables we selected include the closing price, return, volume, one-day lagged volatility of the SZSE 300 Index on a daily basis. Exogenous explanatory variables include the daily return of the SSE 50 index, 3-month treasury bond's daily performance, the daily performance of the 6-month treasury bond, and the daily forex rate of the RMB/USD as inputs for the neural network algorithm.

### B. Algorithm design

On one hand, the advantages of Genetic Algorithm lie in the three following respects[25]. Firstly, it is good at global search so as to avoid local optimal solutions[26-27]. Secondly, it can process multi-modal problems. Also, it is applicable to derivative-free optimization with no need to calculate gradient. On the other hand, neural networks have a strength in processing complicated pattern recognition and forecast problems, extracting features from data training, and timely adjustment for different datasets. Therefore, we can incorporate GA to optimize the structure and parameters of BP network using global search to reduce computing cost, increase training speed and avoid being trapped in local optima.

Specifically, there are three steps to use genetic algorithms to optimize BP neural networks, including choosing the structure of the BP neural network, optimization with genetic algorithm, and prediction.

As Figure 1 below, there are three different layers in a BP neural network. First, we need to determine the number of layers, inputs and output of the fitting formula, and the length of the individual in the genetic algorithm.

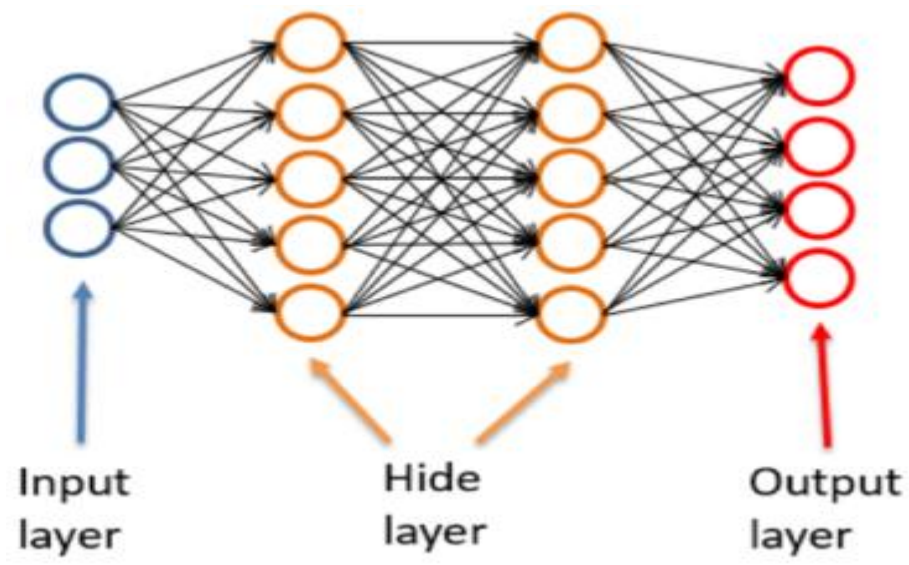

Figure 1. Structure of BP neural network

Then how does genetic algorithm optimize the thresholds and weights of the BP neural network? And every single name in the group contains the weights and thresholds of the entire process. With the adaptive function, each single unit derives its value of fitness, and then, after selection, crossover, and mutation operations, each individual gets its optimal value of fitness. The following Figure 2 displays the evolution process of this approach[28].

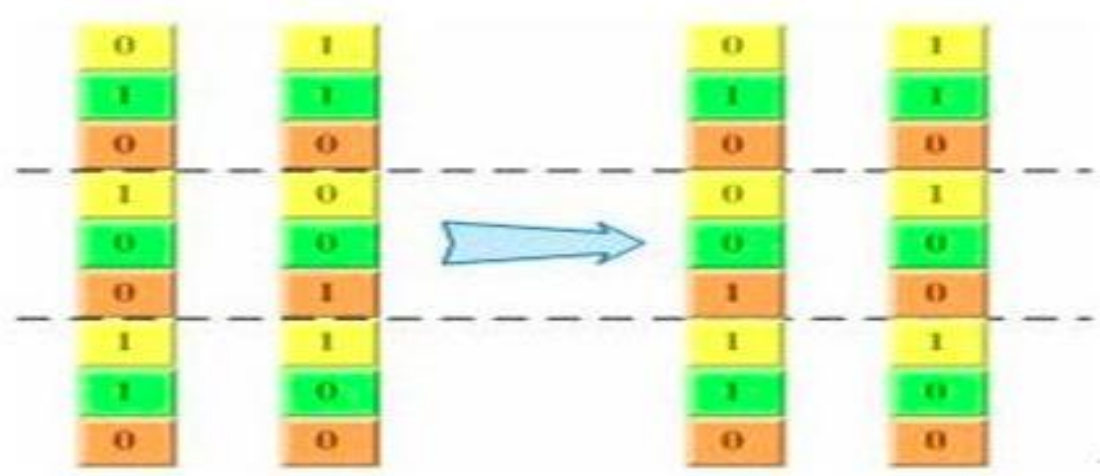

Figure 2. Evolution of GA

Coupled with genetic algorithm, the BP algorithm prediction assigns the initial weights and thresholds of the network to the optimal individual, and the prediction function outputs the network through training. The overall algorithm flow is as following Figure 3:

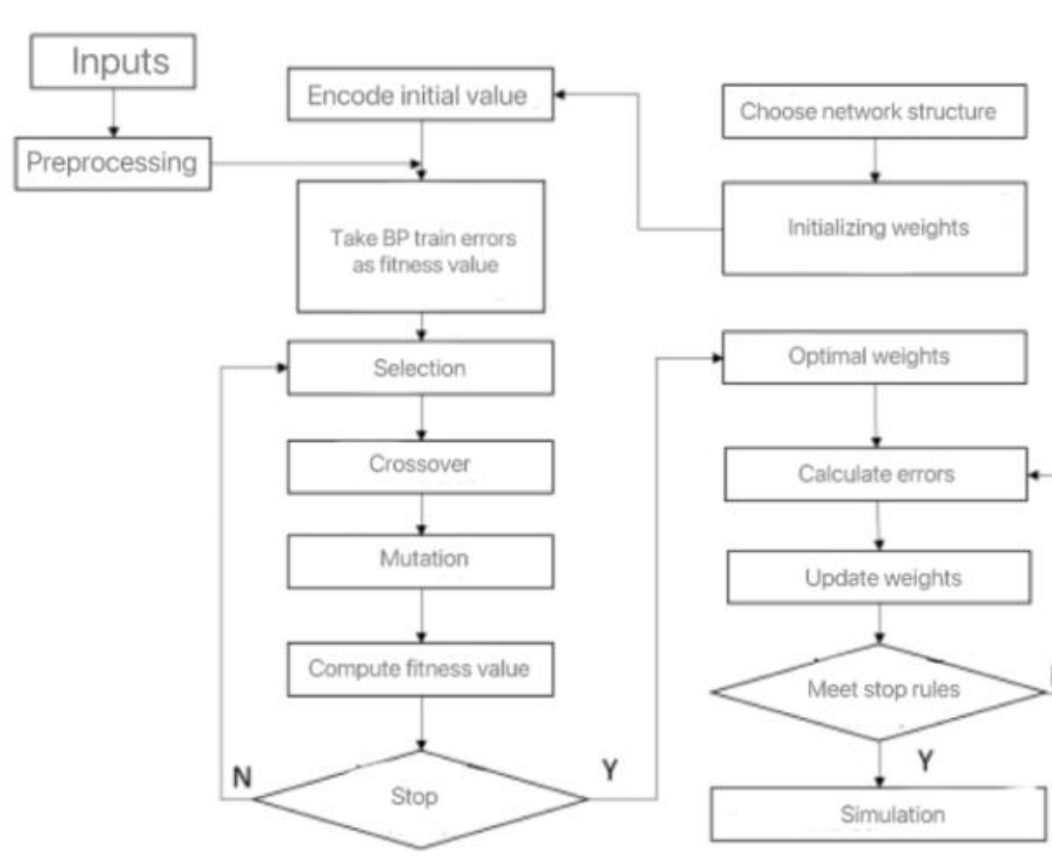

Figure 3. Flow chart of GA-BP neural network

The parts of BP neural network optimized by genetic algorithm optimization contain mutation operation, fitness function, selection operation, crossover operation, and population initialization.

First, the group is initialized. Real number encoding is taken for each member's encoding, where each individual is a string of real numbers, including the hidden layer threshold values, the connection weights between the hidden layer and input layer, the output layer threshold values, and the connection weights among the output layer and the hidden layer. Each member includes all the thresholds and weights of the neural network, and when the network structure is decided, it can generate a neural network by a determined structure, weights, and thresholds.

The second step is construction of fitness function. Generally, use the training datasets error as each member fitness value after training the neural network of BP with training data. Based on the each member, derive the thresholds and initial weights of the BP neural network, train the BP neural network with the training data to make a prediction of the system output, and calculate each single unit's fitness value as $G = k(\sum_{i=1}^{n} abs(y_i - o_i))$, where $y_i$ is the expected output of the BP neural networks, $o_i$ is predicted output of step i, n is the node number in the network output, and k is a parameter[29].

The third step is to select the members with outstanding fitness from the group to form a new population. The selection operation approaches in genetic algorithms are varied, such as tournament method and roulette wheel method. The roulette wheel method, which selects individuals based on the fitness proportion, is applied in this paper. The selection probability of individual i is given by

$$g_i = k/G_i$$
$$p_i = \frac{g_i}{\sum_{j=1}^{N} g_i}$$

In particular, $G_i$ represents the fitness value of member i, and since the value is smaller the better, this value is inverted prior to single name selection. k is the coefficient, and N is the quantity of members in the group.

The fourth step is crossover. A random pair of individuals is selected from the pool, and a new individual is generated by crossing them with a certain probability. Since the individuals are represented by real-number coding, the crossover operation method used is the real-number crossover methodology. The crossover operation method for the $p_{th}$ chromosome $m_p$ and the $r_{th}$ chromosome $m_r$ at position q is as follows:

$$m_{pq} = m_{pq}(1-n) + m_{rq}n$$
$$m_{rq} = m_{rq}(1-n) + m_{pq}n$$

where, n is a number in the range of [0,1].

Step five is mutation. A random member is selected from the pool and a new individual is generated by mutating it with a certain probability. The mutation operation on the $j_{th}$ gene of the $i_{th}$ individual, $m_{ij}$, is as follows:

$$m_{ij} = \begin{cases} m_{ij} + (m_{ij} - m_{max}) * f(g), \ r > 0.5 \\ m_{ij} + (m_{min} - m_{ij}) * f(g), \ r \le 0.5 \end{cases}$$

In this equation, $m_{max}$ represents the upper limit of gene $m_{ij}$, and $m_{min}$ represents the lower limit of $m_{ij}$. The function f(g) is defined as $f(g) = r_2\left(1 - \frac{k}{k_{max}}\right)^2$; where $r_2$ is generated randomly, k is the latest iteration number, $k_{max}$ is the upper limits of evolutionary iterations times, and r is also randomly generated between 0 and 1.

### C. Realized volatility

The realized volatility is used in this paper to assess the accuracy of model predictions. The formula for calculating the realized volatility on the t-th day is as follows:

$$R_Vol_t = \sqrt{\sum_{i=t}^{t+d} (L_i - \bar{L})^2 / d}$$

Where $L_i$ is the logarithmic return, $\bar{L}$ is the average of L, and d is the day number until the nearest option expiration date.

### D. Loss function

The article uses four loss functions, namely mean absolute error (MAE), mean squared error (MSE), root mean squared error (RMSE), and mean absolute percentage error (MAPE), to evaluate the degree to which the model's predicted values deviate from the true values. Usually, with the smaller function value, the model's predicted values are closed to the true values, and the robustness of a model would be better. The calculated formulas are as below:

$$MFE = n^{-1} \sum_{i=1}^{n} (\sigma_i - RV_i)$$

$$RMSE = (n^{-1} \sum_{i=1}^{n} (\sigma_i - RV_i)^2)^{1/2}$$

$$MAE = n^{-1} \sum_{i=1}^{n} |\sigma_i - RV_i|$$

$$MAPE = n^{-1} \sum_{i=1}^{n} |\frac{\sigma_i - RV_i}{RV_i}|$$

## IV. Analysis of Prediction Results

### A. Parametric optimal

The optimal individual fitness values of the genetic algorithm optimization process are displayed in Figure 4:

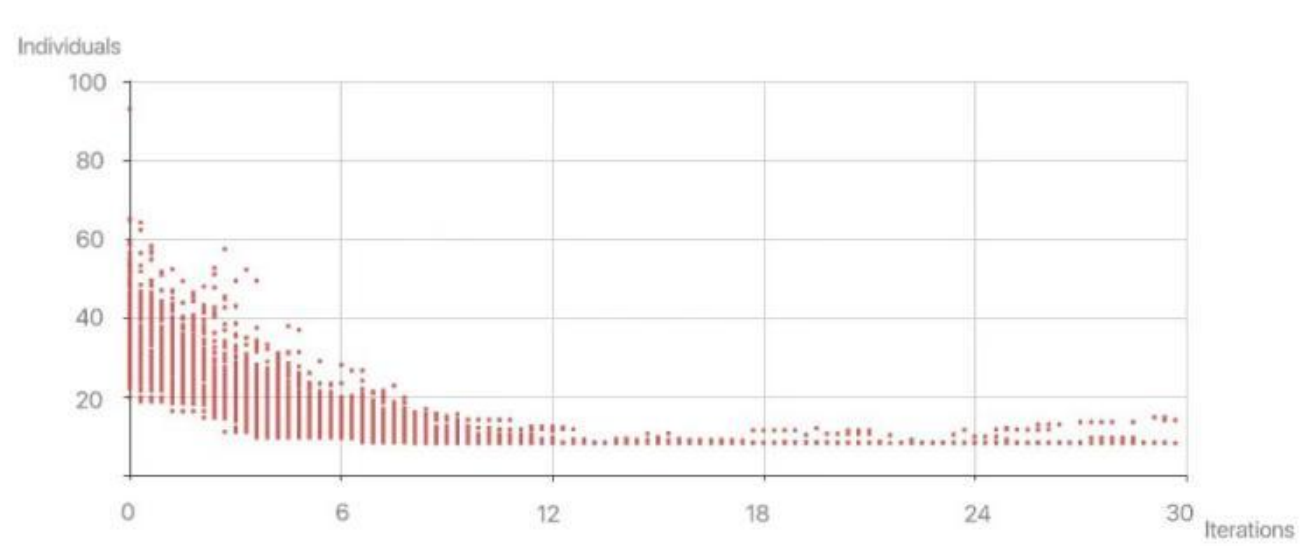

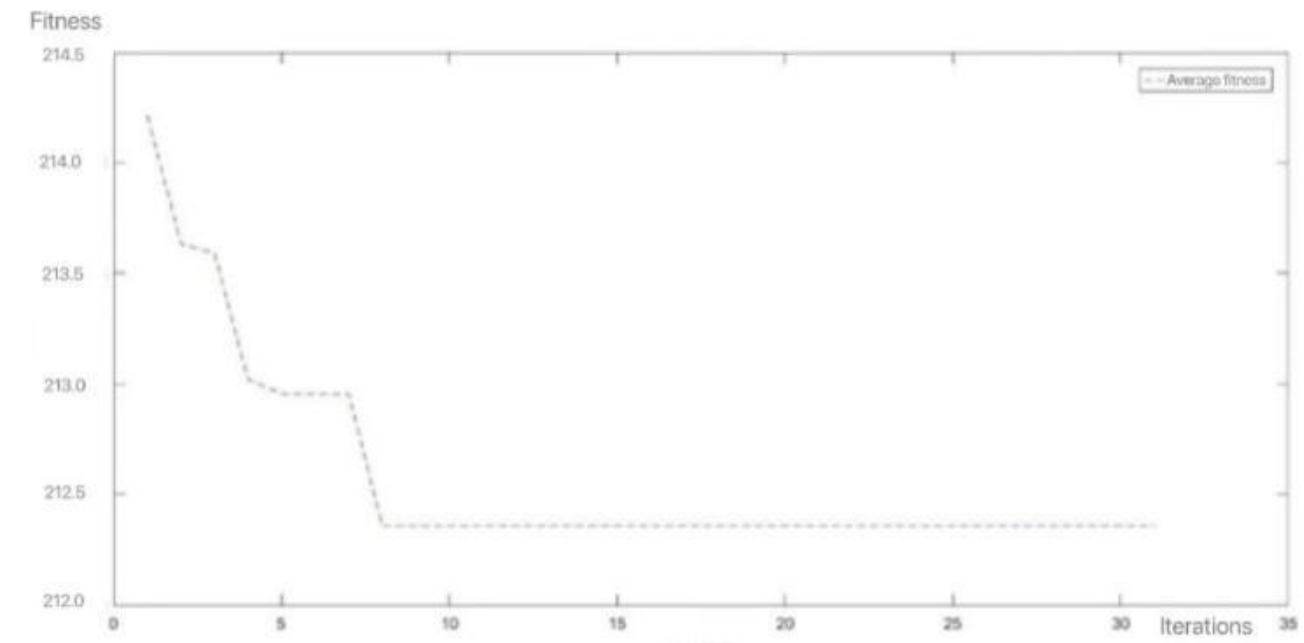

Figure 4. Fitness of optimal individual

The fitness curve terminates at the algebraic expression equals 30. As far as we can see from Figure 4, after iterating 30 times, genetic algorithms can mutate optimal next generations.

### B. Prediction and assessment

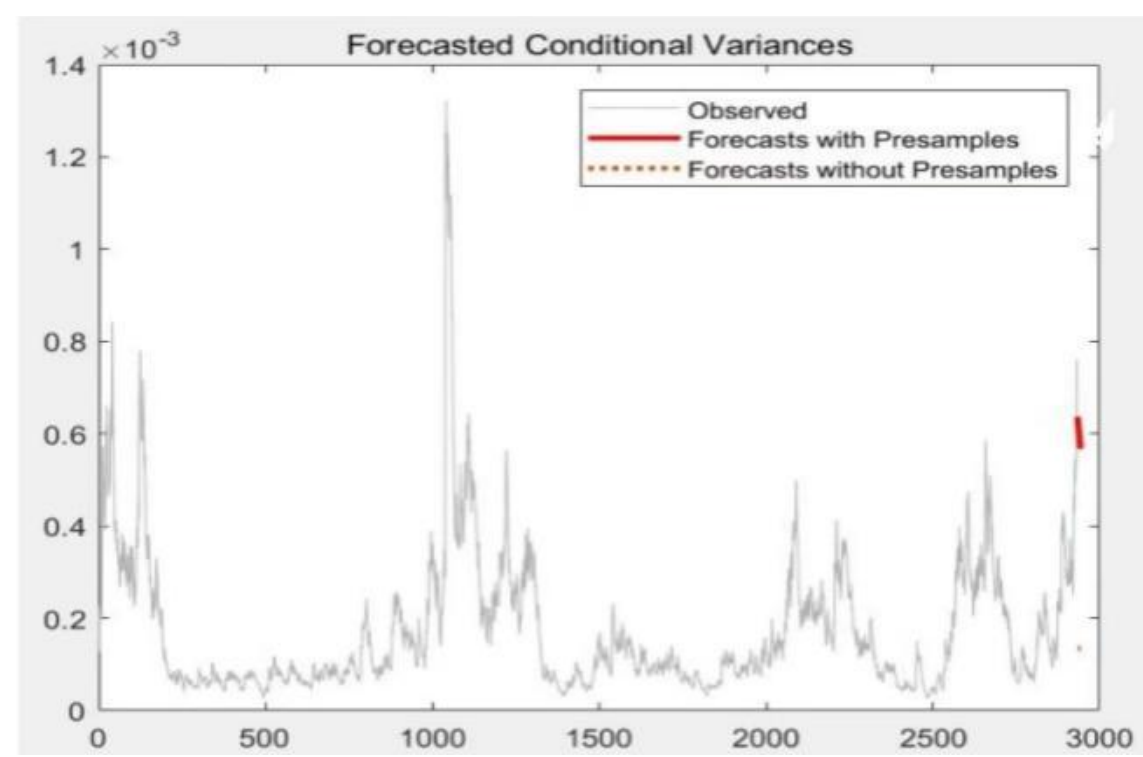

Figure 5. Volatility prediction with GA-BP

Given that datasets are spilt into train set and test set randomly, thus, from Figure 5, we can easily see that, although test set and predicted data is limited here, volatility prediction is performed pretty well.

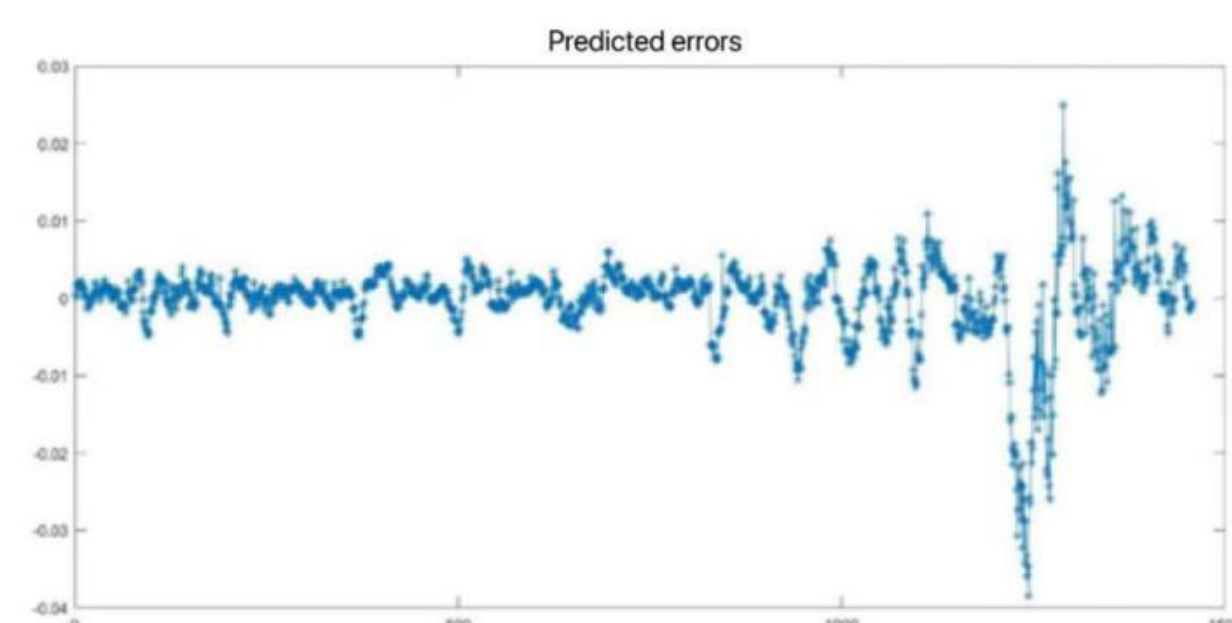

Figure 6. GA-BP neural network predicted errors

From Figure 6, we can see that predicted errors are pretty small, which means, our model predicts pretty well. Of course, predicted errors goes up significantly when run around 1300 times, although it doesn't destroy our model, it makes sense in the future to improve it.

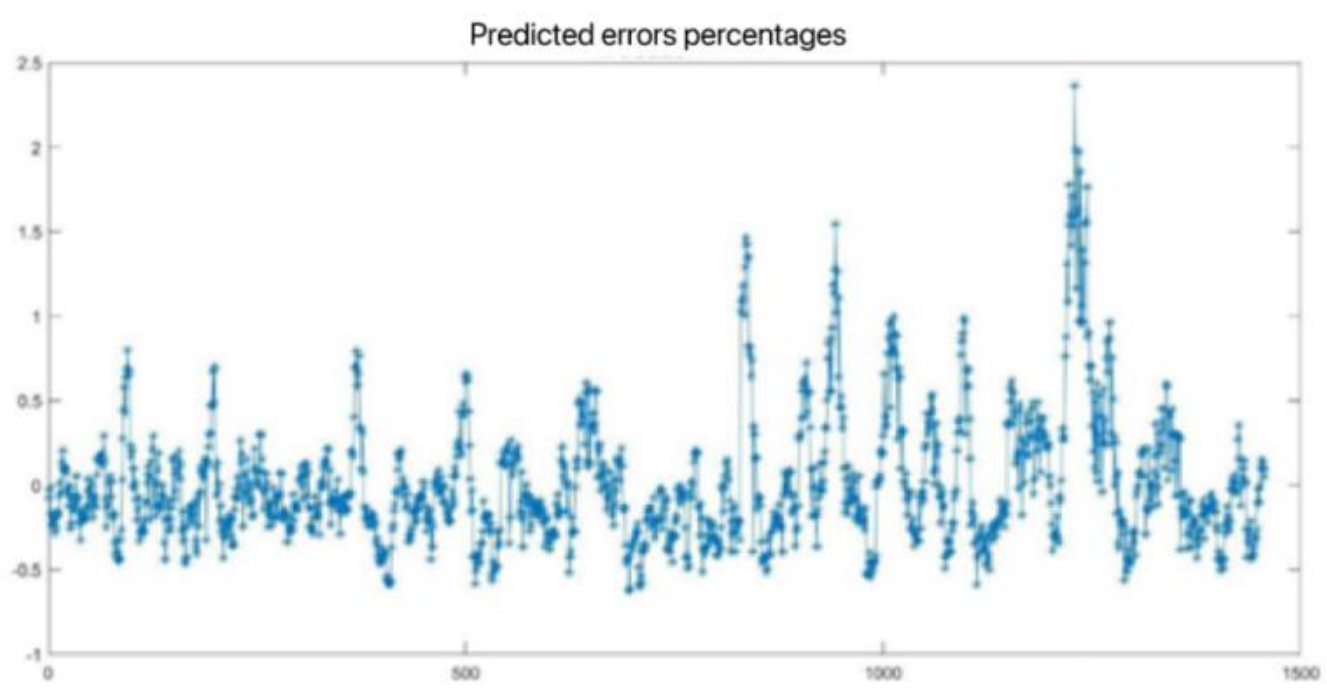

Figure 7. GA-BP neural network predicted errors percentages

Similarly, predicted errors percentages (as Figure 4) also prove that our methodology also works pretty well with extremely low proportion of obvious errors. And during most of time, perhaps, more than 90% of cases, shares of predicted errors are significantly small. Based on our predict charts and error check, it is easy to get a conclusion that, our model is designed reasonable with satisfied prediction ability and low prediction errors.

## V. Conclusions

Based on results above, we can see, that our designed model can predict very well and predicted errors are pretty insignificant. Therefore, genetic algorithms can improve the accuracy of prediction for the original BP neural network. This is due to the GA's strong global search and parallel processing capabilities. Specifically, it can be applied to optimize the learning rule and weights of the neural networks. Moreover, global search and parallel processing enable a higher speed of parameter optimization, and thus accelerate the learning rate of neural networks. Meanwhile, GA can iterate the structure of the neural network to obtain the optimal network structure. Of course, it also performs functional, property, and status analysis of neural networks to help understand the topological structure and its functions.

In the future, some work can be done further. For example, how does it work if incorporating GA-BP with other models, like CNN and random forest? Besides, it would be better to apply the model to other emerging markets, like India, south Africa and Brazil to test its results.